\documentclass[11pt]{article}

\usepackage{moriond_nopic,epsfig,wrapfig}
\usepackage[dvips]{color}
\def\lsim{\raise0.3ex\hbox{$<$\kern-0.75em\raise-1.1ex\hbox{$\sim$}}}
\def\gsim{\raise0.3ex\hbox{$>$\kern-0.75em\raise-1.1ex\hbox{$\sim$}}}
\def\be{\begin{equation}}
\def\ee{\end{equation}}
\def\bea{\begin{eqnarray}}
\def\eea{\end{eqnarray}}
\newcommand{\lspage}[1]   
{ \rotatebox{90}{ \begin{minipage}{\textheight} #1 \end{minipage} } }

\begin{document}

\begin{flushright}
IPPP/03/21\\
DCPT/03/42\\ 
April 2003
\end{flushright}

\vspace*{2cm}

\title{CONSTRAINING THE GLUON IN THE PROTON VIA PHOTOPRODUCTION 
REACTIONS\,\footnote{Talk given at the 
38th Rencontres de Moriond on QCD and Hadronic Interactions, 
Les Arcs, France, March 22--29, 2003.}}

\author{G.~HEINRICH}

\address{Institute for Particle Physics Phenomenology, University of Durham, 
Durham DH1 3LE, UK}

\maketitle
\abstracts{Two reactions, the photoproduction of a direct photon plus a jet 
and the photoproduction  of a charged hadron plus a jet, 
are studied in view of their 
potential to constrain the gluon distribution in the proton. 
The results are based on a program of partonic event generator type 
which includes the full set of NLO corrections.}

\section{Introduction}

The theoretical predictions for prominent processes 
like $gg\to H$ or the $t\,\bar t$ cross section 
still suffer from a rather large uncertainty stemming from 
the parton distributions functions (PDFs) in the proton, 
in particular from the gluon distribution. 
In view of the LHC with its large gluon luminosity, 
it is therefore important to analyse how present experiments 
can be used to constrain the gluon further.
HERA has always played a major role in the task of PDF determinations, 
most importantly through DIS experiments. 
However, the large-$x$ range ($x\,\gsim \,0.2$) 
is still rather poorly constrained, being mainly probed only by
fixed target experiments and recently by the high-$E_T$ 
Tevatron jet data~\cite{highET}.

In photoproduction reactions at HERA, a quasi-real photon, 
emitted at small angle from the electron,  
interacts with a parton from the proton. 
The photon can either participate directly in the hard interaction
or be resolved into a partonic system, in which case a parton
stemming from the photon takes part in the hard interaction. 
Therefore the interest in photoproduction experiments
has rather been focused on measuring the parton distributions 
in the {\it photon} than in the proton. However, 
as will be argued in the following, photoproduction 
reactions can also serve to probe the gluon in the {\it proton}, 
complementary to other measurements.
Here I will consider in particular the following two reactions:
$\gamma \,p \to  \gamma + jet$ and $\gamma \,p \to  h^{\pm} +jet$, 
where $h^{\pm}$ is a charged hadron.  
As compared to dijet photoproduction, these reactions have 
the experimental advantage that photons as well as charged hadrons are 
straightforward to measure. 
However, large-$p_T$ photons can also stem from the fragmentation of a 
hard parton or from light meson decay, such that isolation cuts have to be
imposed. While isolation may introduce a source of systematic errors, 
it also reduces the uncertainty stemming from the photon fragmentation
functions. In the case of charged hadron production, the dependence on the 
hadron fragmentation functions is unavoidable; on the other hand, the
$h^\pm$\,+\,jet cross section has the advantage of being substantially larger
than the $\gamma$\,+\,jet cross section. 

The photoproduction of prompt photons has been measured 
by ZEUS~\cite{Breitweg:2000su} and 
compared to theoretical predictions some time
ago~\cite{Gordon:1995km,Fontannaz:2001ek,Krawczyk}. 
The case where a jet in addition to the prompt photon is also measured 
allows for a more detailed reconstruction of the underlying 
parton dynamics and has recently been analysed in the context of 
a determination of the effective transverse momentum $\langle k_T\rangle$ 
of the partons in the proton~\cite{Chekanov:2001aq,Fontannaz:2001nq}. 
The photoproduction of large-$p_T$  charged hadrons 
has been measured at HERA~\cite{hadrons} 
and compared to theoretical predictions~\cite{Kniehl:2000hk,Fontannaz:2002nu}
only for the inclusive case so far.

\section{Optimisation of the sensitivity to the gluon in the proton}

As observables which serve to reconstruct the longitudinal 
momentum fraction of the parton stemming from the  
proton respectively photon, it is common to use
\be
x_{obs}^{p}=\frac{p_T\,{\rm{e}}^{\eta}+E_T^{\rm{jet}}\,{\rm e}^{
\eta^{\rm{jet}}}}{2E^{p}}\quad,\quad
x_{obs}^{\gamma}=\frac{p_T\,{\rm{e}}^{-\eta}+E_T^{\rm{jet}}\,{\rm e}^{
-\eta^{\rm{jet}}}}{2E^{\gamma}}
\label{xobs}
\ee
where $p_T$ and $\eta$ are transverse momentum and 
pseudo-rapidity\,\footnote{We use the HERA
convention that the proton is moving towards positive rapidities.}
of the final state photon respectively hadron. 
However, as the determination of $E_T^{\rm{jet}}$ introduces a 
source of systematic errors, 
we propose a slightly different variable which does not depend on 
$E_T^{\rm{jet}}$, 
\be
x_{LL}^{p}=\frac{p_T\,(\rm{e}^{ \eta}+\rm{e}^{\eta^{\rm{jet}}})}{2E^{p}}
\quad ,\quad
x_{LL}^{\gamma}=\frac{p_T\,(\rm{e}^{-\eta}+\rm{e}^{-
\eta^{\rm{jet}}})}{2E^{\gamma}}\;.
\label{xll}
\ee
In the case of direct photons in the initial state, the $x_{obs,LL}^{\gamma}$ 
distributions are peaked at values close to one, whereas resolved 
photons correspond to lower values of $x_{obs,LL}^{\gamma}$. 

In order to constrain the gluon distribution function $g(x^p)$ in the proton, 
we need to find a kinematic region where the sensitivity of the cross section 
to $g(x^p)$ is large. Clearly, $g(x^p)$ is large at small $x^p$, which 
means, according to (\ref{xobs}) and (\ref{xll}), at small $\eta,\eta^{\rm{jet}}$. 
On the other hand, small values of  $\eta,\eta^{\rm{jet}}$ imply that 
$x^\gamma$ is large and thus processes initiated by  direct photons dominate
over resolved ones. Therefore, if appropriate rapidity cuts select the backward
region, we expect that the contribution from the gluon in the proton is large, 
while the uncertainty stemming from the parton distributions in the {\it photon}
is minimised, as the resolved photon component is small. 

Let us now verify these considerations by numerical results obtained with the
NLO partonic event generator {\tt EPHOX}\,\cite{Fontannaz:2001nq,Fontannaz:2002nu,phox}. 
We used $\sqrt{s}=300$\,GeV, the MRST01 set\,1 
proton PDFs\,\cite{Martin:2001es}, the AFG photon PDFs\,\cite{afg} and the  
fragmentation functions BFGW\,\cite{Bourhis:2000gs} for the hadrons,  
BFG\,\cite{fragphot} for the photons. 
Further, the cuts $E_T^{\rm{jet}}>5$\,GeV, $p_T^\gamma>6$\,GeV and  $p_T^h>7$\,GeV
have been applied in order to be well within the perturbative region. 

\begin{wrapfigure}[11]{r}[0cm]{6.5cm}
\vspace*{-1.1cm}
\epsfig{file=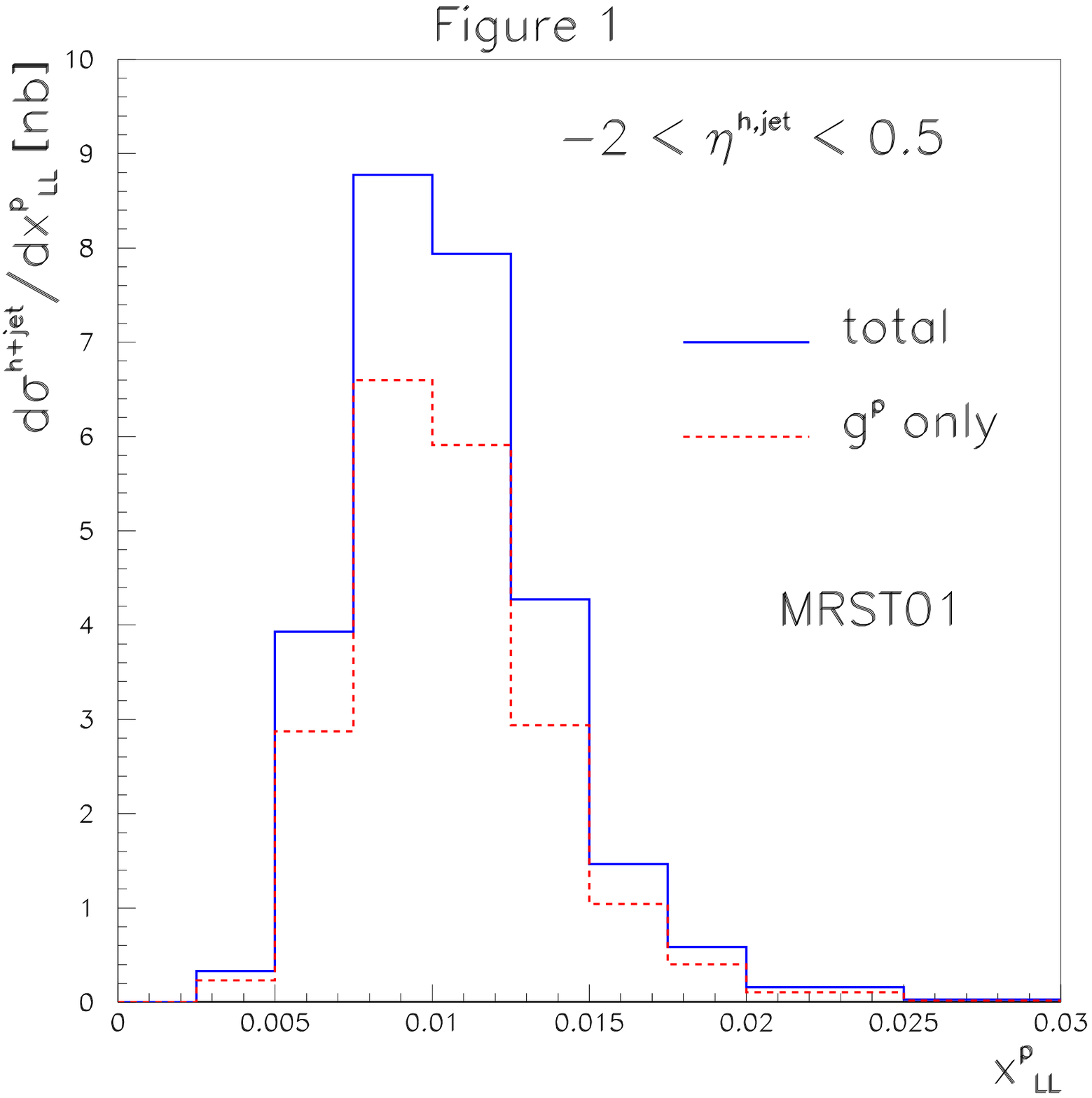, width=6.5cm,angle=0}
\end{wrapfigure}
For the case of hadron+jet production, we can see from Fig.\,1 
that indeed the gluon (from the proton) initiated processes dominate for 
$-2<\eta^{h},\eta^{\rm{jet}}<0.5$.
Unfortunately, the scale dependence 
of the $h^\pm$\,+jet cross section is also large.
Varying the renormalisation scale $\mu$ and the initial/final state 
factorisation scales $M/M_F$  diagonally between $p_T/2$ and $2p_T$ 
leads to a variation of the cross section of about $\pm$20\%. 
On the other hand, a scale optimisation can be 
performed~\cite{Fontannaz:2002nu}, where 
a region of minimal sensitivity can be localised close to 
$\mu=M=M_F=p_T/2$. 

In the case of $\gamma$\,+jet production, the situation is very different. 
As the photon is isolated, the fragmentation contribution to the cross 
section is suppressed, and the leftover candidates for dominant 
subprocesses at small $x^p$ are 
(a) $g^{p}+q^{\gamma},g^{\gamma}\to \gamma+ jet$\, (resolved $\gamma$)\,,
(b) $g^{p}+\gamma\to \gamma+ jet$. 
However, the process (b) only exists at next-to-leading order!
This means that the $g^{p}+\gamma$ initiated subprocess cannot be 
dominant  at small $x^p$ as in the case of $h^\pm$ production. 
Although this looks like an inconvenience, we can turn it into a virtue: 
If we can find a kinematic region where the process (a)
dominates, the sensitivity to 
the gluon in the proton is {\it not} confined to the small $x^p$ range. 
The price to pay is of course a dependence on the photon PDFs, but as 
the quark distributions in the photon are rather well known, the 
task is to isolate a region where the subprocess 
$g^{p}+q^{\gamma}\to \gamma+ jet$ dominates\,\footnote{It has to be stressed
that selecting only particular subprocesses in an NLO calculation is unphysical,
such that the precise magnitude of the subprocess contribution is quite 
sensitive to scale changes. However, it has been verified that the pattern shown
in Fig.\,2 remains valid for other scale choices, as it reflects the 
physical behaviour of the cross section due to the underlying parton dynamics.}. 
\begin{wrapfigure}[20]{r}[1.5cm]{10.9cm}
\vspace*{-1.1cm}
\epsfig{file=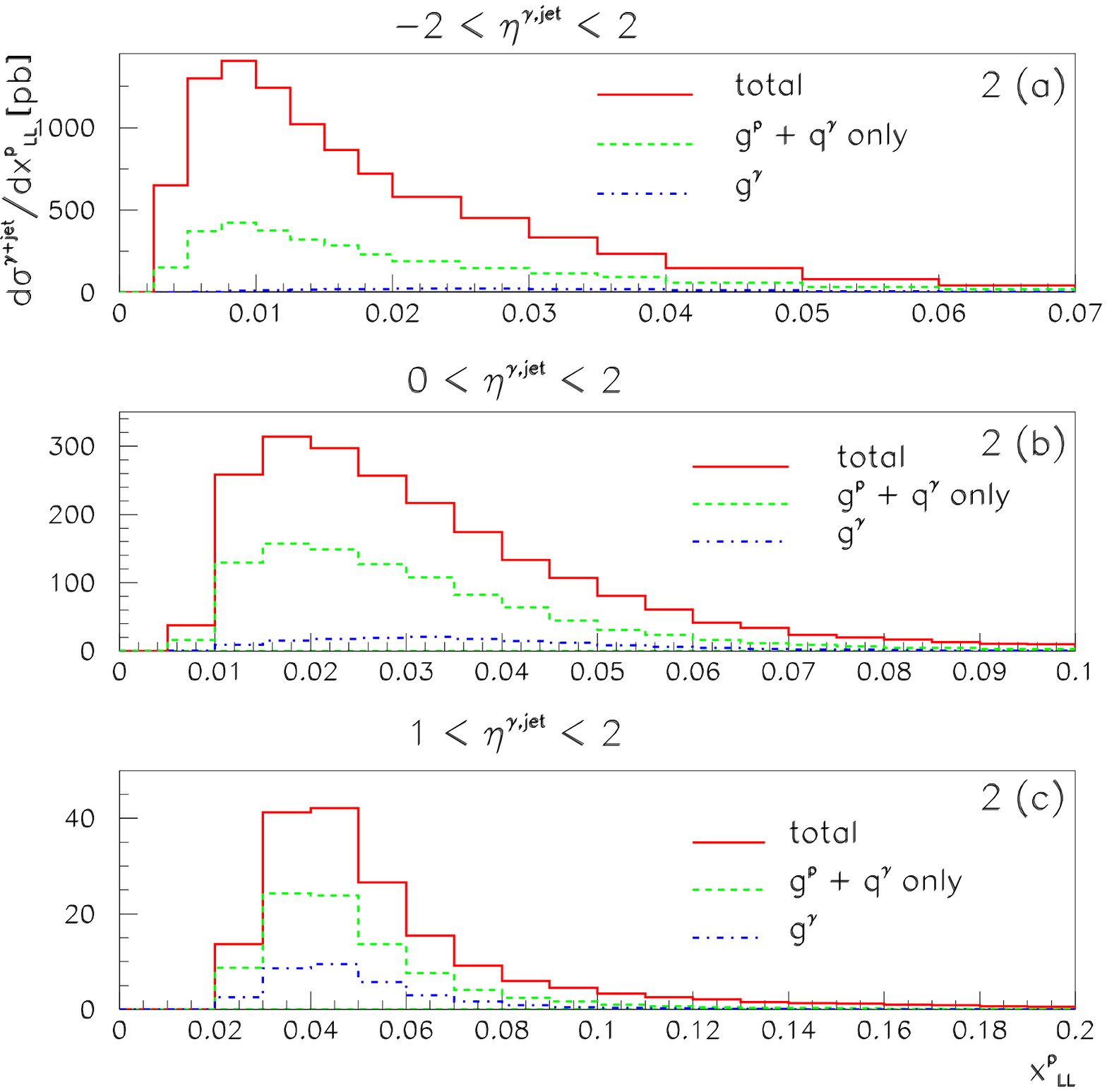, width=10.5cm}
\vspace*{-5mm}
\label{rapcuts}
\end{wrapfigure}
In Fig.\,\ref{rapcuts}
it is shown that this can in fact be achieved by rapidity cuts only. 
Fig.\,\ref{rapcuts}\,(a) shows that the process 
$g^{p}+q^{\gamma}\to \gamma+ jet$ contributes about 30\% to the 
cross section $d\sigma/dx^p_{LL}$ if the whole rapidity range 
$-2<\eta^{\gamma},\eta^{\rm{jet}}<2$ is considered. Selecting the forward region
$0<\eta^{\gamma},\eta^{\rm{jet}}<2$ increases the contribution 
of the $g^{p}+q^\gamma$ initiated processes  to about 50\% of the total, 
while the contribution from the gluon in the photon is still very small, 
as can be seen from Fig.\,\ref{rapcuts}\,(b). If we increase $\eta_{\rm{min}}$ 
even further,  
the total cross section becomes 
rather small and the gluon from the photon starts to become important, 
as shown in Fig.\,\ref{rapcuts}\,(c). Therefore, the region 
$0<\eta^{\gamma},\eta^{\rm{jet}}<2$ is optimal in what concerns the sensitivity 
to $g^{p}$ while minimising the uncertainty from the photon PDFs. 

\begin{minipage}{12cm}
\begin{tabular}{rr}
\includegraphics[width = 5.9cm]{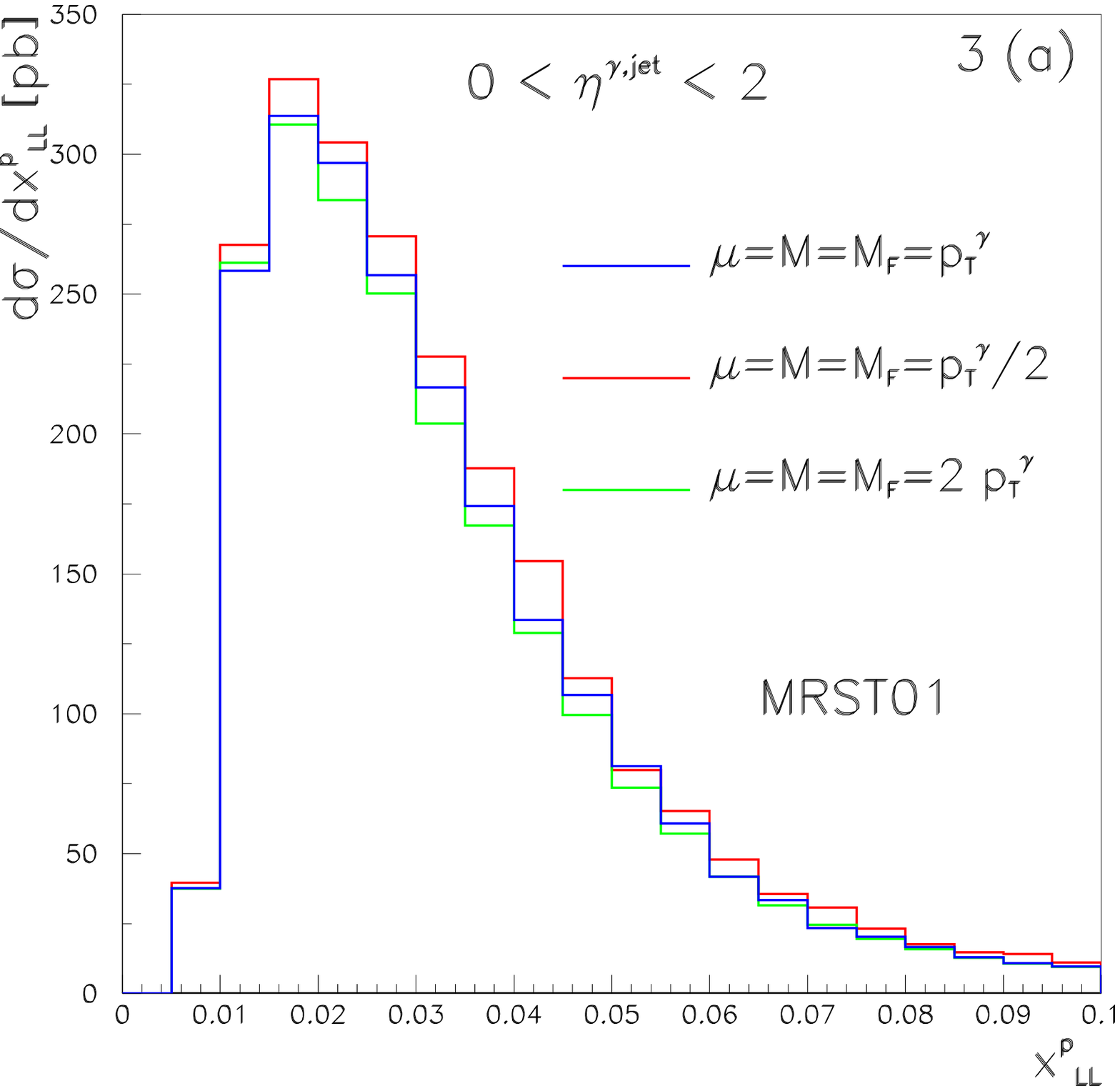} &
\begin{minipage}{6cm}
\vspace*{-5.7cm}
\includegraphics[width = 5.9cm]{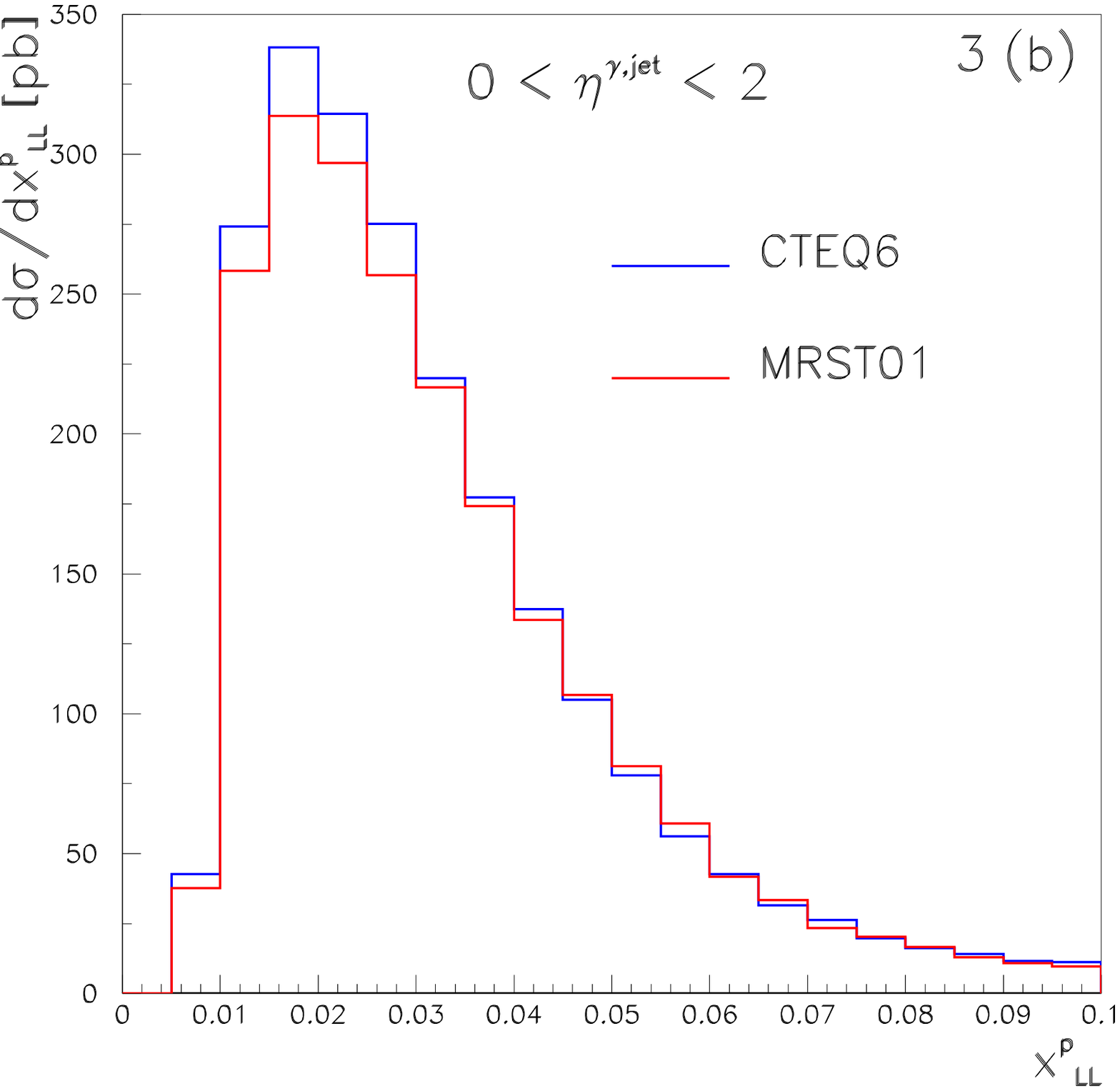}
\end{minipage}
\end{tabular}
\end{minipage}

An important feature of the $\gamma+jet$ cross section is 
that the scale dependence is 
very weak. From Figs.\,3\,(a),(b) one can see that in the bins around 
0.02, where the difference between the MRST01 and CTEQ6 parametrisations is
significant, the uncertainty due to scale variations is smaller than the
difference between these two parametrisations.

\begin{wrapfigure}[13]{r}[0.5cm]{6cm}
\vspace*{-1.2cm}
\epsfig{file=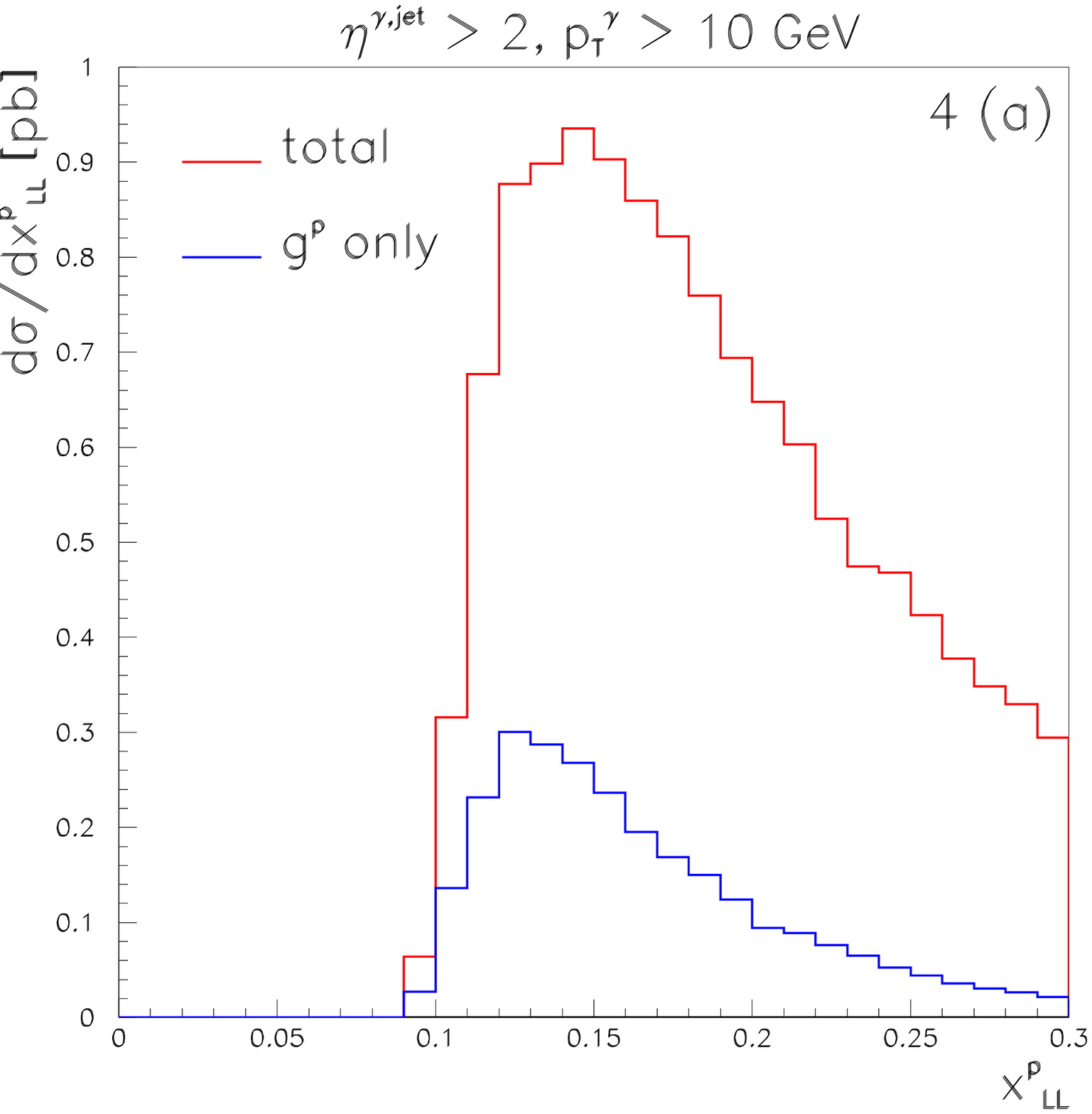,height=6.2cm}
\end{wrapfigure}

Nevertheless, the region $0<\eta^{\gamma},\eta^{\rm{jet}}<2$ 
does not select the high-$x$ range. 
In order to access $x$ values beyond $x^p\,\gsim \,0.1$, more 
stringent cuts have to be placed, selecting the very forward region, 
which has not been accessible by HERA \nolinebreak 1, but will be accessible by 
future HERA experiments. 
As an example, it is shown in Fig.\,4\,(a)
that $x^p_{LL}$ values $> 0.1$ can be achieved by the cuts 
$\eta^{\gamma},\eta^{\rm{jet}}>2,\, p_T^{\gamma}>10$\,GeV. 
\begin{wrapfigure}[8]{r}[0.5cm]{6cm}
\vspace*{-2.7cm}
\epsfig{file=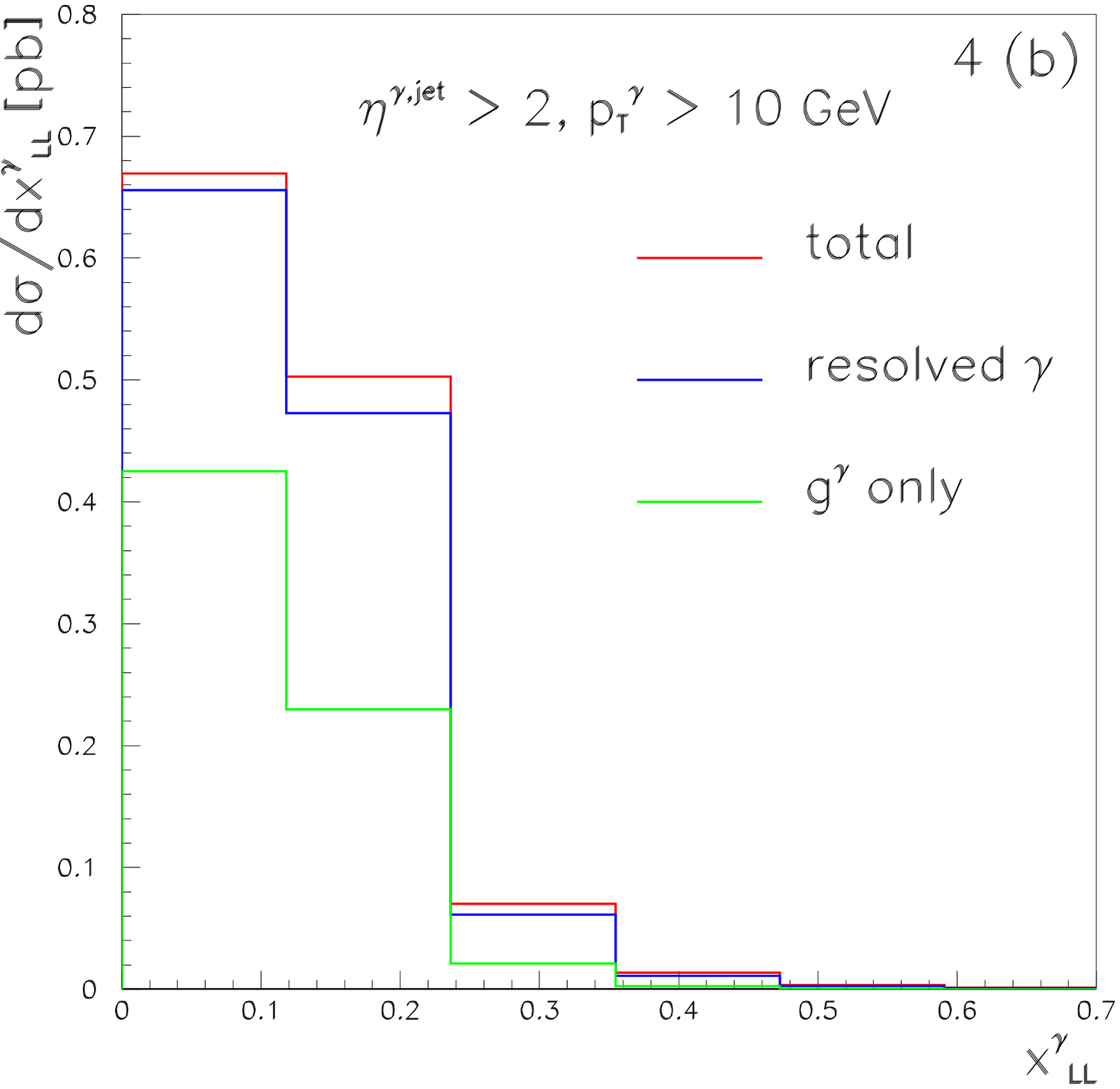,height=6.2cm}
\end{wrapfigure}
However, the 
contribution from the gluon in the proton is rather small in this region. 
On the other hand, the contribution from 
the gluon in the {\it photon} is substantial, such that  this region 
is favourable to pin down the gluon distribution in the  photon, 
as can be seen from Fig.\,4\,(b).

\section{Conclusions and outlook}
The $h^\pm+jet$ cross section  shows a large sensitivity to the gluon distribution 
$g(x^p)$ in the proton around $x^p\sim 0.01$, but it suffers from a 
scale dependence of the order of 20\%. 
In contrast, the scale dependence of the $\gamma + jet$ cross section 
is quite weak ($\lsim\,7$\%).  
In the $\gamma + jet$ case, maximising the sensitivity to $g(x^p)$ while at the same 
time minimising 
the uncertainty from the photon PDFs suggests the rapidity cuts 
$0<\eta^{\gamma},\eta^{\rm{jet}}<2$, probing the $x^p$-range  
$0.01 \,\lsim\,x^p\,\lsim \,0.05$. 
Future HERA experiments with an improved forward tracking system and high
statistics could allow to cut even more towards forward rapidities 
and thus access  $x^p$ values of about 
$0.1 \,\lsim\,x^p\,\lsim \,0.3$.

\vspace*{5mm}
\noindent{\bf Acknowledgements}\\
I am grateful to my collaborators M.~Fontannaz and J.-Ph.~Guillet.  
I also would like to thank the organisers of the Moriond conference.

\end{document}